\documentclass[12pt,a4paper,twocolumn]{narms2}

\usepackage{subfigure} % was: subfignarms.sty
\usepackage{psfig}
\usepackage{timesmt}   % use times typeface
\usepackage{chikako}   % <---------------------------------- MOD
\usepackage{graphicx} \usepackage{subfigure}
\usepackage{amssymb}
\makeatletter
\renewcommand{\section}{\@startsection%
{section}%
{1}%
{0mm}%
{- \baselineskip}%
{0.15\baselineskip}%
{\normalfont\normalsize}}%

\renewcommand{\subsection}{\@startsection
{subsection}%
{2}%
{0mm}%
{-\baselineskip}%
{0.15\baselineskip}%
{\normalfont\normalsize}}%
\makeatother

\begin{document}
%\maketitle

\title{Response of a hexagonal granular packing under a localized
overload: effects of pressure} \author{\large {S. Ostojic*}\\ {\em
Institute for Theoretical Physics, Universiteit van Amsterdam,
Valckenierstraat 65, 1018 XE Amsterdam,} \\ {\em The Netherlands}\\
\\
\large {D. Panja}\\ {\em Institute for Theoretical Physics,
Universiteit Utrecht, Postbus 80.195, Leuvenlaan 4, 3508 TD Utrecht,}
\\ {\em  The Netherlands}\\ } \date{}% No date.

\abstract{We study the response of a two-dimensional hexagonal packing of rigid,
frictionless spherical grains due to a vertically downward point force
on a single grain at the top layer. We use a statistical approach,
where each configuration of the contact forces is equally likely. We
find that the response is double-peaked, independantly of the details
of boundary conditions. The two peaks lie precisely
on the downward lattice directions emanating from the point of
application of the force. We examine the influence of the confining
pressure applied to the packing.}

%%%%%%%%%%%%%%%%%%%%%%%%%%%%%%%%%%%%%%%%%%%%%%%%%%%%%%%%%%%%%%%%%%%

%  To kick in the changes.
% Make title here for two column format.
\maketitle
\frenchspacing   % no double spaces after colon
                 % added by <W.Hennings@fz-juelich.de>

% Write some ascii text files called intro.tex, concept.tex, etc.
% TeX and LaTeX will look for the .tex subscript by default.

\section{INTRODUCTION}

Force transmissions in (static) granular packings have attracted a lot
of attention in recent years
\shortcite{jaeger:rev,bouch:rev,fluct:mueth,fluct:blair1,vanel,qmodel,ned,resp:gold1,resp:geng1,resp:geng2}. 
Granular
packings are assemblies of macroscopic particles that interact only
via mechanical repulsion effected through physical
contacts. Experimental and numerical studies of these systems have
identified two main characteristics. First, large fluctuations are
found to occur in the magnitudes of inter-grain forces, implying that
the probability distribution of the force magnitudes is rather broad
\shortcite{fluct:mueth,fluct:blair1}. Secondly, the average propagation of forces
--- studied via the response to a single external force --- is
strongly dependent on the underlying contact geometry
\shortcite{vanel,resp:geng1,resp:geng2}.

Most of the available theoretical models capture either  one or the
other of these two aspects. The scalar $q$-model \shortcite{qmodel}
reproduces the observed force distribution reasonably well, but yields diffusive
propagation of forces, in conflict with experiments
\shortcite{resp:geng1}. Similarly, continuum elastic and elasto-plastic
theories \shortcite{ned} have been used  by engineers for years, but
they
%D provide  a description only at the average macroscopic level. 
provide no information on the distribution of force magnitudes.  More
ad-hoc ``stress-only'' models \shortcite{bouch:rev} include structural
randomness, but its consequences on the distribution of forces are
unclear.
 
A simple conjecture, which provides a fundamental principle for the
study of  both fluctuations and propagation of forces,  has been put
forward by Edwards years ago \shortcite{edwards1}. The idea
is to consider all ``jammed'' configurations equally likely. {\it A
priori}, there is no justification for this ergodic hypothesis, but
its application  to models of jamming and compaction has been rather
successful \shortcite{jam:makse1}. Its extension to
forces in granular packings is in principle straightforward: sets of
forces belonging to all mechanically stable configurations have equal
probability. However, in an ensemble of stable granular packings, two
levels of randomness are generally present
\shortcite{bouch:rev}. First, the force geometry clearly depends on
the underlying geometrical  contact network, which is different in
different packings. Secondly, randomness in the values of the forces
is present even in a fixed contact network, since the forces are not
necessarily uniquely determined from the contact network. Instead of
considering both levels of randomness simultaneously, a natural first
step is thus to obtain the averages for a fixed contact geometry, and
then possibly  average over the contact geometries.

\begin{figure}[h]
\begin{center} 
\includegraphics[width=0.9\linewidth]{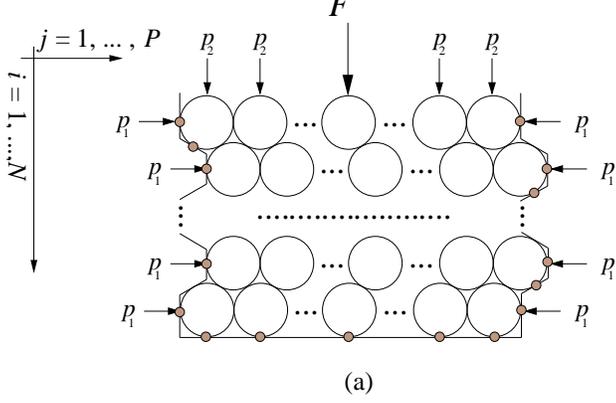}
\caption{Our model: $N\times P$ array of hexagonally close-packed
rigid frictionless spherical grains in two-dimensions (drawn for odd
$N$). At the top, a uniform vertical pressure $p_2$ is applied on all
grains, and an overload $F-p_2$ is added on one grain. On the sides, a
uniform pressure $p_1$ is applied on the packing (little gray circles
appear on interfaces where the contact forces are non-zero). }
\label{fig1}
\end{center}
\end{figure}

Such  a method  has been  shown  to produce  single inter-grain  force
probability  distributions in  fixed geometry  that compare  well with
experiments  \shortcite{fluct:jacco1}.  Following  a similar  path, we
studied the response of  a two-dimensional hexagonal packing of rigid,
frictionless  spherical  grains (see  Fig.\ref{fig1})  to a  pointlike
external force  $F$, and found that  it is concentrated  mainly on the
two lattice directions emanating from  the point of application of the
force    \shortcite{jstat},     in    agreement    with    experiments
\shortcite{resp:geng1}. Our definition of  the response of the packing
is         $G(i,j)=\left[\langle        W_{i,j}\rangle        -\langle
W_{i,j}^{(0)}\rangle\right]/F$ where $W_{i,j}$ and $W_{i,j}^{(0)}$ are
the vertical  force transmitted  by the $(i,j)$th  grain to  the layer
below it respectively with and  without the external overload $F$. The
angular  brackets denote  averaging  with equal  probability over  all
configurations of repulsive contact forces in mechanical equilbrium.

Below we summarize the method and the results
%D , and study in more detail 
regarding the effects of
%D the 
confining pressure. More precisely, we study two cases: (i) $p_2=0$;
(ii) $p_1=p_2\equiv p$.

\section{CONCEPTUAL MODEL}
%hint: for citations according to Balkema style, use "\shortcite{key}"
%      instead of "\cite{key}

To start with, we describe a method for assigning the uniform
probability measure on the ensemble ${\mathcal E}$ of {\it stable
repulsive contact forces\/} pertaining to a fixed geometrical
configuration of $P$ {\it rigid, frictionless two-dimensional disks of
arbitrary  radii\/}. The directions of the forces are fixed at each of
the $Q$ contact points, and one can represent any force configuration
by a column vector $\mathsf F$ consisting of $Q$ non-negative
force magnitudes
$\{F_{k} \}$ (with ${k=1,\ldots, Q}$) as its individual
elements. These elements satisfy $2P$ Newton's equations, which can be
represented as ${\mathbf A}\cdot{\mathsf F}={\mathsf
F}_{\mbox{\scriptsize ext}}$. Here, $\mathbf{A}$ is a $2P\times Q$
matrix, and $\mathsf{F}_{\mbox{\scriptsize ext}}$ is a
$2P$-dimensional column vector representing the external forces.  If
we assume $2P<Q$ --- as in the hexagonal packing we consider --- then
there is no unique solution for $\mathsf{F}$. In that case, there
exists  a whole set of solutions that can be constructed  via the
three following steps: (1)  one first identifies an orthonormal basis
$\{{\mathsf F}^{(l)}\}$ ($l=1,\ldots,d_K=Q-2P$) that spans the space
of $Ker({\mathbf A})$; (2) one then determines a unique solution
${\mathsf F}^{(0)}$ of ${\mathbf A}\cdot{\mathsf F}^{(0)}={\mathsf
F}_{\mbox{\scriptsize ext}}$  by requiring ${\mathsf F}^{(0)}.{\mathsf
F}^{(l)}=0$ for $l=1,\ldots,d_K$; and (3) one finally obtains all
solutions of ${\mathbf A}\cdot{\mathsf F}={\mathsf
F}_{\mbox{\scriptsize ext}}$  as ${\mathsf F}={\mathsf
F}^{(0)}+\sum\limits_{l=1}^{Q-2P}f_l\,{\mathsf F}^{(l)}$, where $f_l$,
for $l=1 \ldots Q-2P$,
are real numbers. This implies that  ${\mathcal E}$ is parametrized by
the $f_l$'s belonging to a set ${\mathcal S}$ obeying the
non-negativity conditions  for all forces, i.e. ${\mathsf
F}^{(0)}_{k}+\sum\limits_{l=1}^{Q-2P}f_l\,{\mathsf F}^{(l)}_{k}>0$,
$\forall k=1\ldots Q$ . The uniform measure on
${\mathcal E}$  is thus
equivalent to the uniform measure $d\mu=\prod\limits_{k} dF_k\,
\delta({\mathbf A}\cdot{\mathsf F}-{\mathsf F_{\mbox{\scriptsize
ext}}} )\,\Theta({F_k})=\prod\limits_{l} df_l$ on ${\mathcal S}$.

\section{ANALYSIS}
\begin{figure}[h]
\begin{center} 
\includegraphics[width=0.95\linewidth]{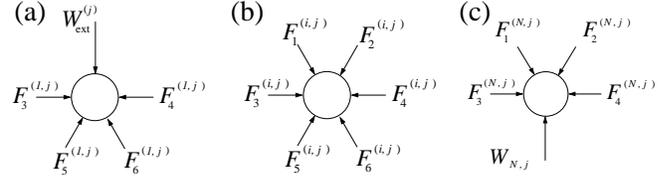}
\caption{Schematically shown forces on the $j$th grain in the $i$th
layer: (a)$i=0$, $W_{ext}^{(j)}=p_2+\delta_{j,j_0}F$ (b) $i\leq N$,
(c) $i=N$; $F^{(i,j)}_m\geq0\,\,\forall m$.}
\label{fig2}
\end{center}
\end{figure}
We will now apply the general method described above  to a hexagonal
packing of monodisperse, massless, rigid and  frictionless disks
subject to uniform confining pressures $p_1$ from the sides and $p_2$
from the top (see. Fig. \ref{fig1}). Our aim is to calculate the mean
of $W_{i,j}=\frac{\sqrt{3}}{2}\left[F^{(i,j)}_1+F^{(i,j)}_2\right]$
(see Fig. \ref{fig2}) when an overload $F$ is applied on the top of
the packing.

Force balance on individual grains (see Fig.  \ref{fig2}) can be
expressed as
\begin{eqnarray}
  F^{(1,j)}_{5}=\frac{1}{\sqrt{3}}W_{\mathrm{ext}}^{j}+[F^{(1,j)}_4-F^{(1,j)}_3]\nonumber\\
 F^{(1,j)}_{6}=\frac{1}{\sqrt{3}}W_{\mathrm{ext}}^{j}-[F^{(1,j)}_4-F^{(1,j)}_3]\nonumber\\
F^{(i,j)}_{5}=F^{(i,j)}_{2}+[F^{(i,j)}_4-F^{(i,j)}_3]\nonumber\\
F^{(i,j)}_{6}=F^{(i,j)}_{1}-[F^{(i,j)}_4-F^{(i,j)}_3]\nonumber\\
W_{N,j}=\sqrt{3}\left[F^{(N,j)}_1+F^{(N,j)}_2\right]/2\nonumber\\
F^{(N,j)}_4-F^{(N,j)}_3=\left[F^{(N,j)}_1-F^{(N,j)}_2\right]/2\,.
\label{force_balance}
\end{eqnarray}

Note that these equations can be solved {\it grain by grain} for
increasing $i$ and $j$ starting from the top layer. For $i=1$ and
$j=1$, the top and side forces are known from the boundary conditions:
$W_{\mathrm{ext}}^{1}=p_2$ and $F_3^{(1,1)}=p_1$ . Then, if
$F_4^{(1,1)}$ is chosen randomly, $F_5^{(1,1)}$ and $F_6^{(1,1)}$ are
uniquely determined from (\ref{force_balance}). Similarly, for the
next grain $i=1$ and $j=2$, once $F_4^{(1,2)}$ is fixed, $F_5^{(1,2)}$
and $F_6^{(1,2)}$ are uniquely determined from
$W_{\mathrm{ext}}^{2}=p_2$, and $F_3^{(1,2)}=F_4^{(1,1)}$. This
procedure can be carried on identically for all the other grains for
$i=1$, except for $j=P$ where $F_4^{(0,P)}$ is set by the boundary
condition. In this way, the top forces $F_1^{(i,j)}$ and $F_2^{(i,j)}$
on all grains in the layer $i=2$ are known, and Eq.
(\ref{force_balance}) can be solved in a similar fashion to determine
the forces for all layers successively.

Such a sequential procedure shows that, once the forces $F^{(i,j)}_4$
are fixed, all the others are uniquely determined. In other words
$F^{(i,j)}_4$'s for $1\leq i< N,1\leq j< P$  parameterize ${\mathcal
E}$. The number of these parameters is indeed $d_K$, as it should
be. Clearly, in this formulation
$d\mu=\prod\limits_{(i,j)=(1,1)}^{(N-1,P)}\!\!dF^{(i,j)}_4$ with
$F_4^{(i,j)}>0$.  Moreover to respect the non-negativity conditions
for $F^{(i,j)}_5$'s and $F^{(i,j)}_6$'s, the parameters
$F_4^{(i,j)}$  must satisfy
\begin{eqnarray}
\displaystyle{-F^{(i,j)}_2\leq F_4^{i,j}-F_3^{i,j} \leq F^{(i,j)}_1}\,,
\label{Gineq}
\end{eqnarray}
implying that the set ${\mathcal S}$ of allowed values of
$F_4^{(i,j)}$'s is bounded and the uniform measure on ${\mathcal S}$ 
well-defined.

\section{$q$-COORDINATES AND COMPUTATIONAL SCHEME} 

%D To
In order to evaluate $\langle W_{i,j} \rangle= \displaystyle{\frac{1}{{\mathcal
N}}\int}_{\!\!\mathcal S} \,W_{i,j}\prod_{kl} dF^{(k,l)}_4$
where ${\mathcal N}=\int_{{\mathcal S}}\prod_{ij}dF_4^{i,j}$ is the
normalization constant, we define
\begin{eqnarray}
q_{i,j}=\left[\sqrt{3}(F_4^{(i,j)}-F_3^{(i,j)}+F^{(i,j)}_2)/2\right]/W_{i,j}\,,
\label{varchange}
\end{eqnarray}
where $q_{i,j}$ is the fraction of $W_{i,j}$ that the $(i,j)$th grain
transmits to the layer below it towards the left, i.e., $F^ {(i,j)}_5=
2q_{i,j} W_{i,j}/\sqrt{3}$ and $F^{(i,j)}_6=2(1-q_{i,j})
W_{i,j}/\sqrt{3}$.  Clearly, $W_{0,j}$ are the external forces applied
on the top layer. For $i>0$, $W_{i,j}$ is a function of $q_{k,l}$ for
$k<i$, since
\begin{eqnarray}
W_{i,j}=(1-q_{i-1,j-1})\,W_{i-1,j-1}+q_{i-1,j}\,W_{i-1,j}.
\label{qprop}
\end{eqnarray}

The probability measure on the new coordinates $q_{i,j}$ is given by
the Jacobian of the variable change (\ref{varchange})
\begin{equation}
\textstyle{\prod\limits_{i,j}}\,dF_4^{(i,j)}=2^{NP}\,\textstyle{\prod\limits_{i,j}}
dq_{i,j}\,W_{i,j}(q)\big/3^{NP/2}\,.
\label{jac}
\end{equation}
%D This is the 
There is a fundamental difference between the model we study and
the $q$-model \shortcite{qmodel}: while in the $q$-model the $q$'s
corresponding to different grains are 
%D usually 
chosen to be
uncorrelated and identically distributed, the uniform  measure on
${\mathcal S}$ translates to the measure $\prod_{ij} W_{i,j}(q)$, which is
not factorizable in a product of identical factors depending on a
single $q$.

The non-negativity conditions (\ref{Gineq}) for $F^{(i,j)}_5$ and
$F^{(i,j)}_6$ are automatically satisfied for $0 \leq q_{i,j} \leq
1$. However, the conditions $F_4^{(i,j)}\geq 0$ must be taken into
account separately; 
%D further restricting $q_{i,j}$ between $0$ and $1$.
they may further restrict the choice of available values of the
$q_{i,j}$'s. 

Using the $q$-coordinates, the $\langle W_{i,j}\rangle$ can be evaluated
numerically with a variant of the Metropolis algorithm \cite{newbark},
the usual Boltzmann factor being replaced by $\prod_{ij} W_{i,j}$. At each
Monte Carlo step, a grain is chosen randomly, and the corresponding
value of $q$ is drawn uniformally between $0$ and $1$. The values of
$W_{i,j}$ and of the lateral forces $F_4^{(i,j)}$ are evaluated on the
affected grains. If any $F_4^{(i,j)}$ becomes negative, the move is
rejected, otherwise the new configuration is accepted with the usual
Metropolis acceptance ratio.

\section{RESULTS}

\begin{figure}
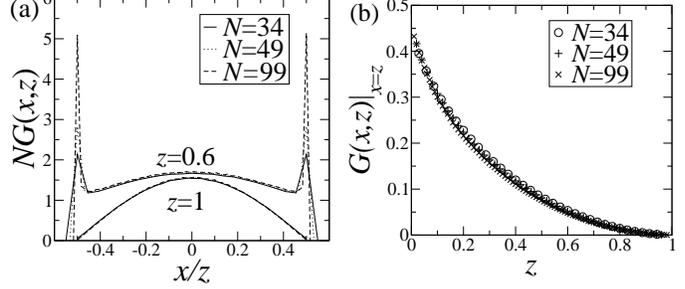

\begin{center}
\begin{minipage}{0.49\linewidth}
\includegraphics[width=0.97\linewidth]{fig4}
\end{minipage}
\begin{minipage}{0.49\linewidth}
\includegraphics[width=0.99\linewidth]{fig5}
\end{minipage}
\end{center}
\caption{Simulation results for $p_2=0$ and $p_1\geq
\frac{2}{\sqrt{3}}F$. Behavior of  $G$ in reduced
co-ordinates $x$ and $z$: (a) scaling of $G(x,z)$ with
system size for $|x|<z/2$ at two $z$ values; (b) data collapse for
$G(x,z)|_{x=z/2}$ at three $N$ values. \label{weightless}}
\end{figure}

We first consider $p_2=0$. In that case, the $q_{i,j}$'s are allowed
to take {\it all\/} values between $0$ and $1$ if
$F/p_1\leq\frac{\sqrt{3}}{2}$. The response $G(i,j)$ then scales
linearly with $F$ (hence we use $F=1$),  and $G(i,j)\equiv0$ outside
the triangle formed by the two downward lattice directions emanating
from $(0,j_0)$, the point of application of $F$.

Our simulation results for $G(i,j)$ within the triangle appear in
Fig. \ref{weightless}. We find $\forall N$ that (i) the $G(i,j)$
values display two {\it single-grain-diameter-wide symmetric peaks\/}
that lie precisely on the boundary of the triangle, (ii) the
magnitudes of these peaks decrease with depth, and a broader central
peak starts to appear, (iii) only a central maximum for $G(i,j)$
remains at the very bottom layer ($i=N$).

We now define $x=(j-j_0)/N$ and $(j-j_0+1/2)/N$ respectively for even
and odd $i$, and $z=i/N$ (see Fig. \ref{fig1}) in order to put the
vertices of the triangle formed by the locations of non-zero $G(i,j)$ values on $(0,0), (-1/2,1)$ and $(1/2,1)\,\,\forall
N$. We denote by $G(x,z)$ the response as function of these rescaled
coordinates.
The excellent data collapse indicates that the $G(x,z)$ values for $|x|<z/2$ scale with the inverse system
size [Fig. \ref{weightless}(a); we however show only two $z$ values],
while the $G(x,z)$ values for $|x|=z/2$ lie on the
same curve for all system sizes [Fig. \ref{weightless}(b)]. The data
suggest that in the thermodynamic limit $N\rightarrow\infty$, the
response $G(x,z)$ scales $\sim 1/N$ for $|x|<z/2$, but
reaches a {\it non-zero\/} limiting value on $|x|=z/2\,\,\forall
z<1$. We thus expect $\lim\limits_{N\rightarrow\infty}
G(x,z)\big|_{|x|=z/2}> G(x,z)\big|_{|x|<z/2}\,\,\forall z<1$; or equivalently, a {\it
double-peaked response at all depths $z<1$ in the thermodynamic limit}.

For $F/p_1>\frac{\sqrt{3}}{2}$ and $p_2=0$, as the value of $p_1$
decreases, the $q_{i,j}$'s get restricted to narrower ranges within
$(0,1)$ with increasing values of $F/p_1$, and consequently, an
increasing amount of vertically downward force carried by the grains
is transferred to the boundary of the triangle. In the limit $F/p_1
\to\infty$, the non-negativity conditions of all the forces make
all $F_3^{(i,j)}$ and $F_4^{(i,j)}$ vanish, and the packing effectively
becomes rectangular. In fact, the same behaviour also
occurs for any value of $p_2\neq0$.

\begin{figure}
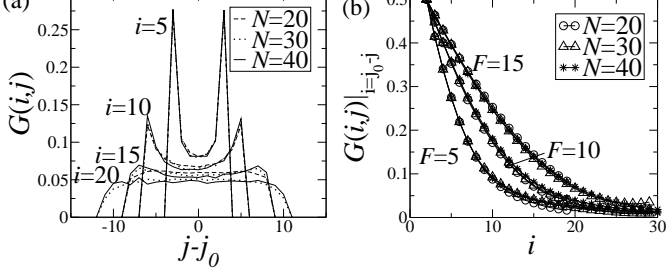

\begin{center}
\begin{minipage}{0.49\linewidth}
\includegraphics[width=0.98\linewidth]{fig6}
\end{minipage}
\begin{minipage}{0.49\linewidth}
\includegraphics[width=0.98\linewidth]{fig7}
\end{minipage}
\end{center}
\caption{Simulation results for $G(i,j)$, for $p_1=p_2=1$: (a)
  $G(i,j)$ for $F=10$ as function of $j-j_0$, at four different values
  of $i$ for various system sizes; (b) $G(i,j)$ as function of $i$ for
  $j_0-j=i$, for $F/p=5,10$ and $15$ and three different system sizes.
\label{pressure}}
\end{figure}

For $p_2\neq0$ however, experimentally the most relevant case is
$p_1=p_2=p$, since in practice a confining pressure has to be applied
on the packing. In this case, the response clearly depends on the
dimensionless parameter $F/p$. The simulation results for $N=20, 30$
and $40$ are presented in Fig. \ref{pressure}.

Fig. \ref{pressure} (a) shows the response for $F/p=10$ as function of
$j-j_0$ for various values of $i$. For small $i$, the response
displays two single-grain-diameter-wide symmetric peaks along the
lattice directions emanating from the point of application of the
overload. As $i$ increases, the magnitude of the peaks decreases, and
the response for large $i$ becomes essentially flat. The dependance of
the response on the system size in this case however is very different
from the case $p_2=0$. Rather then being self-similar, the response is
independant of $N$: $G(i,j)$ at a given depth $i$ is the same for all
systems with $N>i$ layers.

The values of $G(i,j)$ on the lattice direction $i=j_0-j$ as function
of $i$ are plotted in Fig. \ref{pressure}(b), for $N=20, 30$ and $40$,
and $F/p=5, 10$ and $15$. For increasing $F/p$, the values
$G(i,j_0-i)$ increase for fixed $i$ and their decay with increasing
$i$ is slower, implying that the peaks are more pronounced and
propagate deeper in the packing.

\section{CONCLUSIONS}

In summary, we find that assigning equal probability to all
mechanically stable force configurations for rigid, frictionless
spherical grains  in a two-dimensional hexagonally close-packed
geometry yields a double-peaked response independantly of the details
of the boundary condition. The peaks are single grain diameter wide,
and they lie on the two downward lattice directions emanating from the
point of application of $F$. In the case of zero top pressure, the
response exhibits self-similarity, but when the packing is confined by
uniform pressure, the peaks penetrate the packing deeper with larger
$F$. Such a simple model is in good qualitative agreement with
experiments in a reasonably robust manner. Whether grains with
friction \shortcite{clem:fric} produce broadening of the peaks or not
remains to be investigated.

\bibliographystyle{chikako}    % <---------------------------------- MOD
\bibliography{gran_stat_short} % 

\end{document}